\documentclass[]{aastex}
\usepackage{emulateapj5,apjfonts}

\slugcomment{To appear in Sept.\ 10, 2002 ApJ}
\shorttitle{X-ray/H$\alpha$ filaments in NGC 3079}
\shortauthors{Cecil, Bland-Hawthorn, \& Veilleux}

\begin{document}

\title{Tightly Correlated X-ray/H$\alpha$ Emitting 
Filaments In the Superbubble and Large-Scale Superwind of NGC 3079}

\author{Gerald Cecil}
\affil{Department of Physics and Astronomy, University of North Carolina,
Chapel Hill, NC 27599-3255}
\email{gerald@thececils.org}

\author{Joss Bland-Hawthorn}
\affil{Anglo-Australian Observatory, Epping, NSW, Australia}
\email{jbh@aao.gov.au}

\and{}

\author{Sylvain Veilleux\altaffilmark{1}}
\altaffiltext{1}{Cottrell Scholar of the Research Corporation}
\affil{Department of Astronomy, University of Maryland, College Park, MD 20742}
\email{veilleux@astro.umd.edu}

\begin{abstract}
Using \emph{Chandra} and \emph{HST}
we show that X-ray and H\( \alpha  \) filaments
that form the 1.3-kpc diameter superbubble of NGC 3079 have strikingly similar
patterns at $\sim$0\farcs8 resolution. This tight optical line/X-ray match
seems to arise from cool disk gas that has been driven by the wind,
with X-rays being emitted from
upstream, stand-off bowshocks or by conductive
cooling at the cloud/wind interfaces.
We find that the soft X-ray plasma has thermal and kinetic energies
$E_\mathrm{TH} \sim 2\times 10^{56} \eta_x^{0.5}$ erg and 
$E_\mathrm{KE} \sim 5\times10^{54} \eta_x^{0.5}$ erg, 
where $\eta_x$ is the filling factor of the X-ray gas and may be small;
these are comparable to the energies of the optical line-emitting gas.
Hydrodynamical simulations reproduce the observations well using large
filling factors for both gas phases; assuming otherwise leads to
serious underestimates of the mass lost in superwinds and 
therefore their influence within and around the host galaxy.
X-rays are also seen from the base of
the radio counterbubble that is obscured optically by the galaxy disk, and
from the nucleus (whose spectrum shows the Fe K$\alpha$ line at 6
keV as well as gas absorbed by a moderate neutral hydrogen column).
The superbubble is surrounded by
a fainter conical halo of X-ray emission that fills the area delineated
by high angle, H\( \alpha  \)-emitting filaments, supporting our
previous assertion that these filaments form the contact discontinuity/shock
between galaxy gas and shocked wind. This X-ray emission is not significantly
edge brightened, indicating a partially filled volume of warm gas
within the shocked wind, not a shell of conductively heated gas. About
40\arcsec\ (3 kpc) above the galaxy disk an X-ray arc may partially close
beyond the bubble, but the north-east quadrant remains open at the surface
brightness attained by \emph{Chandra}, consistent with the
notion that the superwind has broken out into at least the galaxy halo. 
\end{abstract}


\section{Introduction}

Winds are believed to play a key role in determining the equation of state
of the intergalactic medium and the long-term evolution of galaxies, yet
direct proof is elusive. Wind
energetics have proved very difficult to constrain because of uncertain gas
filling factors in the different phases, but evidence is growing 
that they may be far larger than inferred from optical spectra of the
shock or photoionized nebulae, and radio observations of supernovas and their
remnants.
Winds can influence galaxy evolution in many ways \citep{he02}. 
They are important
redistributors of metals by radiation on dust or by over-pressured gas from
supernova heating. In the Galaxy, enrichment of the halo and outer disk through
powerful winds may account for the observed abundances of the thick disk and 
globular clusters \citep{bf02}. Moreover, winds can regulate the growth of
the central bulge \citep{ca99}, and alter the evolution of satellites
of large galaxies, with either galaxy hosting the outflow
\citep[for example]{ir87}.
Winds may mix astrated gas into intergalactic clouds \citep{tr02},
and may clear paths
through dusty veils for AGN/starbursts to reionize the universe.

A superwind is driven by the pressure of thermalized supernova or AGN ejecta.
As the wind expands and cools adiabatically, it accelerates to several
thousand km~s\( ^{-1} \). Expanding along the pressure gradient in
the galaxy, the wind can over-run, crush, and heat disk and halo clouds
to X-ray emitting temperatures, then entrain
them into the flow. X-rays may also arise from stand-off bow shocks
in the wind upstream of the optically emitting clouds. Thus, soft X-rays may
point to powerful shocks. However, they can also be emitted at the conductive
interface between hot shocked wind and cooler ISM swept up by a bubble. 
Whether gas blows out of the galaxy depends on hydrodynamical details and 
halo drag \citep{st01}.

To understand this ubiquitous phenomenon, we are studying energetic
outflows in nearby galaxies. Here, increased spatial resolution of
many ISM phases may uncover evidence for previous outflows and energetic
stellar populations, constraining the duty-cycle of ejection and illuminating
the feedback role of fast and slow winds on supernovae \citep{ef00}. 
One of the clearest examples of a nearby
superwind-blown structure is the prominent bubble \citep{fo86}
that protrudes from
the nucleus of the nearly edge-on SB(s)c galaxy NGC 3079 (\( 17.3\pm 1.5 \)
Mpc, so 1\arcsec\ = 84 pc). Rising 1.3 kpc (15\arcsec) above
the disk, the bubble is the optically emitting half of
a double-lobed radio structure \citep{du88}, optical line
emission from the other lobe being extinguished by the galaxy disk.
Outflow kinetic energies of up to $5\times10^{55}$ erg are
implied by the optical line emission \citep{ve94}.
Because of considerable extinction, the nuclear power source is manifest
primarily by reprocessed radiation. How the superbubble is powered
is thus uncertain, with both AGN \citep{ha95} and starburst \citep{so01}
sources promoted. 

The total energy required corresponds to
the effects of 10\( ^{4} \)/$\varepsilon$ supernovae, where
$\varepsilon$ is the thermalization efficiency \citep{cc85}.  A
particular obstacle to achieving realistic simulations of the wind is that
even a rough estimate of $\varepsilon$ is controversial \citep{st01}:
it can be $\ll10\%$ in dense gas \citep{th98}
where the energy is radiated away, but can exceed 30\% if the
supernovae occur in multi-phase gas.

In \citet[CBVF hereafter]{ce01}, we analyzed high-resolution images
of the superbubble obtained with \emph{HST}/WFPC2, see Fig. \ref{fig1}a.
The H\( \alpha + \){[}N II{]}\( \lambda 6583 \) emission-line image
shows many plausibly windblown features, including: 1) linear filaments
rising several kpc above the disk to form an ``X''
that we have interpreted to be the contact discontinuity/ISM shock
between wind and galaxy {}``thick disk''/halo gas, 2) four towers
of intertwined filaments that together form the optical emitting walls
of the superbubble and that all disperse at the same height above
the disk. The latter pattern, together with 3) a break in the velocity
field at this point, argue for a vortex in the top half of the bubble
caused by fluid instabilities, as predicted from numerical simulations
of windblown bubbles. 4) Line of sight velocities that increase with height
above the disk to reach 1000 km~s\( ^{-1} \) relative to systemic
at the vortex, decreasing
thereafter, 5) a circumnuclear disk region {}``scoured''
of dust and gas whose boundary joins the linear filaments 1).

X-rays could arise when the H\( \alpha  \) filaments
moving at the speeds seen in the superbubble collide with stationary gas, 
and indeed \citet[PTV hereafter]{pi98}
found that NGC 3079 is ten times more X-ray luminous than galaxies
of similar optical luminosity.  They used the \emph{ROSAT} HRI to image
X-rays across the superbubble with integral 0.1-2.4 keV luminosity
\( 10^{40} \) erg s\( ^{-1} \), and also used 
the \emph{ROSAT} PSPC to detect the softest X-rays out to 2\farcm5
east of the nucleus and 6\arcmin\ west. 
No spectrum could be isolated from the superbubble because
the HRI was energy insensitive while the PSPC lacked
spatial resolution to connect optical and X-ray emitting filaments. 
Now a \emph{Chandra} X-ray image
is available, and the striking correspondence it reveals and the implications
on soft X-ray emission from superwind bubbles are the subjects of
this paper. In \S2 we introduce the X-ray image, and analyze it in
\S3. In \S4 we discuss the dynamics, and summarize in \S5.

\section{Observations and Reductions}

We analyzed the archived 26.9 ksec exposure (P.I. D. Strickland) obtained
with the ACIS-S3 detector on the \emph{Chandra} X-ray Observatory;
the other ACIS-S chips miss the galaxy so are not discussed in this
Letter. We extracted a 2\arcmin\ square region centered on the superbubble
(the aim point of the observation); well exposed point sources are
round in this region, confirming the excellent spacecraft aspect solution.
A few background {}``flares'' were removed
by using the \emph{ciao2.2.1} \textbf{lightcurve} task, reducing
the effective exposure to 26.0 ksec. We formed several images with
different smoothing, using constant and adaptive kernel sizes. For study of
the nucleus and superbubble we obtained best results using a fixed
kernel of 0\farcs8 FWHM. Elsewhere, we mapped fainter emission using
a kernel of 2\arcsec\ FWHM. 

Fig. \ref{fig1}b registers the X-ray image with the \emph{HST} line
image discussed in CBVF. As others have found in recent studies, the
``world coordinates'' derived from calibrations of both spacecraft
agree within 0\farcs7, which is adequate given the smoothing required
for this fairly shallow \emph{Chandra} exposure. We confirmed registration
to this accuracy using centroids of a few compact galaxies visible
in both wavebands.

To form spectra (Fig.\ \ref{fig0}),
we subtracted the background summed from regions off the galaxy to successfully
remove spectral artifacts, and
mapped pulse heights to energies using the \( 32\times 32 \) response 
values at the bubble position in CALDB file acisD2000-01-29fef\_phaN0001.
After subtracting the background, 1757 X-ray events were detected within a
27$\times$22\arcsec\ ellipse centered on the superbubble (Fig.\ \ref{fig0}b)
and excluding a $5\farcs9\times3\farcs9$ ellipse that spans the
nucleus and jet and that emitted 309 events (Fig.\ \ref{fig0}a). 
We used the \emph{Sherpa} package within the \emph{ciao 2.2.1} software
to quantify the spectra and to integrate fluxes.
We fit the spectrum of the bubble with single temperature,
collisional-ionization-equilibrium models such as \texttt{mekal} 
and external photoelectric absorption, allowing neutral absorption,
gas temperature, and interpolated gas 
abundances to vary over the fitting energy interval
$0.15-2.5$ keV; abundances relative to He were fixed
to the \citet{an89} ratios. For the nucleus+jet spectrum, we
also included a single Gaussian at the Fe emission.

Fit results, including
derived X-ray fluxes and luminosities, are summarized in Table 1.
We find that iron and $\alpha$-element abundances
are $\sim15\pm5\%$ solar across the bubble 
($\sim8$ times larger than \citet{st02} derived for the 
diffuse halo of NGC 253), and about half solar near the nucleus.
The lack of independent constraints on
element abundances from optical spectra added uncertainty on the fitted
emission ``integral'' $EI = n_e n_H \eta_x V$, where $\eta_x$ is the X-ray 
volume filling factor. It is possible that depletion onto dust
explains the low iron abundances in both regions, but this would not explain
depleted $\alpha$-elements if confirmed by deeper spectra.
Calibration uncertainties (see \S2.1.4 in \citet{st02}, notably the
effective area of chip S3 is uncertain below 0.5 keV)
and low counts from both regions limited
possibilities for detailed spectral analysis; in particular, the necessary
spectral binning evident in Fig.\ \ref{fig0}
prevented us from using the intensity ratios of successive oxygen
ionization states (VII/VIII) to check the overall temperature solution hence
the questionably low fitting abundance of the bubble.
Low counts also prohibited multi-temperature models. 
We did not explore the addition of a non-thermal component to increase 
gas abundances in the thermal gas although this is clearly plausible physically
not only near the nucleus (where it might be the AGN), but also in the
superbubble (where it might be synchrotron emission from cosmic rays
accelerated locally in shocks).

\section{Physical Properties of Regions Emitting X-rays}

We now discuss the X-ray emission, proceeding from smallest to largest
scale. We do not discuss the considerable number of X-ray sources
evident in the galaxy disk beyond 20\arcsec\ radius.

\subsection{Extended Nucleus}

The nucleus is obscured optically but prominent to \emph{Chandra}.
X-ray flux is elongated along the axis of the radio jet (\S3.2).
The nominal registration puts the nucleus 0\farcs4
south-west of the X-ray flux centroid, of order our registration uncertainty.
Indeed our spectral model shows that flux near the 
nucleus is absorbed $\sim10\times$ more than
\citet{pt97} estimated from the combined \emph{ASCA+ROSAT} PSPC spectrum;
we could not fit the \emph{Chandra} spectrum with his prefered absorption.
Fig.\ \ref{fig0}a shows relatively strong Fe K$\alpha$ near 6 keV, 
providing unambiguous evidence for the AGN
that generally excites this line by fluorescence of the cold molecular gas
evident nearby \citep{so01a}.
Sparse counts in the adjacent ``continuum''
prevented us from establishing the equivalent width of this line to compare
to its strength in other AGN.

There is no evidence in the spectrum for a hard-component due either to the
AGN or to unrelated hard-spectrum X-ray binaries down to a count rate of 
0.004 per second integrated over the interval $3-7$ keV.

\subsection{Optical Extension of the VLBA-Scale Jet}

Panels A of Fig.\ \ref{fig1}c show \emph{anticorrelated} X-ray and
high-velocity optical line filaments associated with the VLBI-scale
jet (CBVF). The main optical filament (top one
in panel A) is displaced from an X-ray source (top oval in the
top panel A) that does not show the hard spectrum of an
unrelated X-ray binary down to the aforementioned count rate.
The X-rays are likely not
inverse Compton scattered because there is no strong IR emission
or collimated radio flux
along this axis at the scale of the H$\alpha$/X-ray emission.
Instead, the thermal spectrum that we infer with $kT\sim 0.5$ keV is 
consistent with that expected for a jet cocoon shock with average expansion or
entrainment velocity 
$V_{s7}=\sqrt{T_s x_{ts}/3.2\times 10^5} = 600$ km~s$^{-1}$,
where $V_{s7}=V_{shock}/100$ km~s$^{-1}$, and
$x_{ts}=2.3$ is the concentration of all particles relative to hydrogen
nuclei when the preshock medium is fully ionized and has 10\% helium abundance
by number \citep{hm79}. This velocity is consistent with measured emission-line
half-widths in the region, see Figs.\ 11b \& c of CBVF.

\subsection{\label{sec:towers}Bubble {}``Towers''}

\begin{figure}
\includegraphics{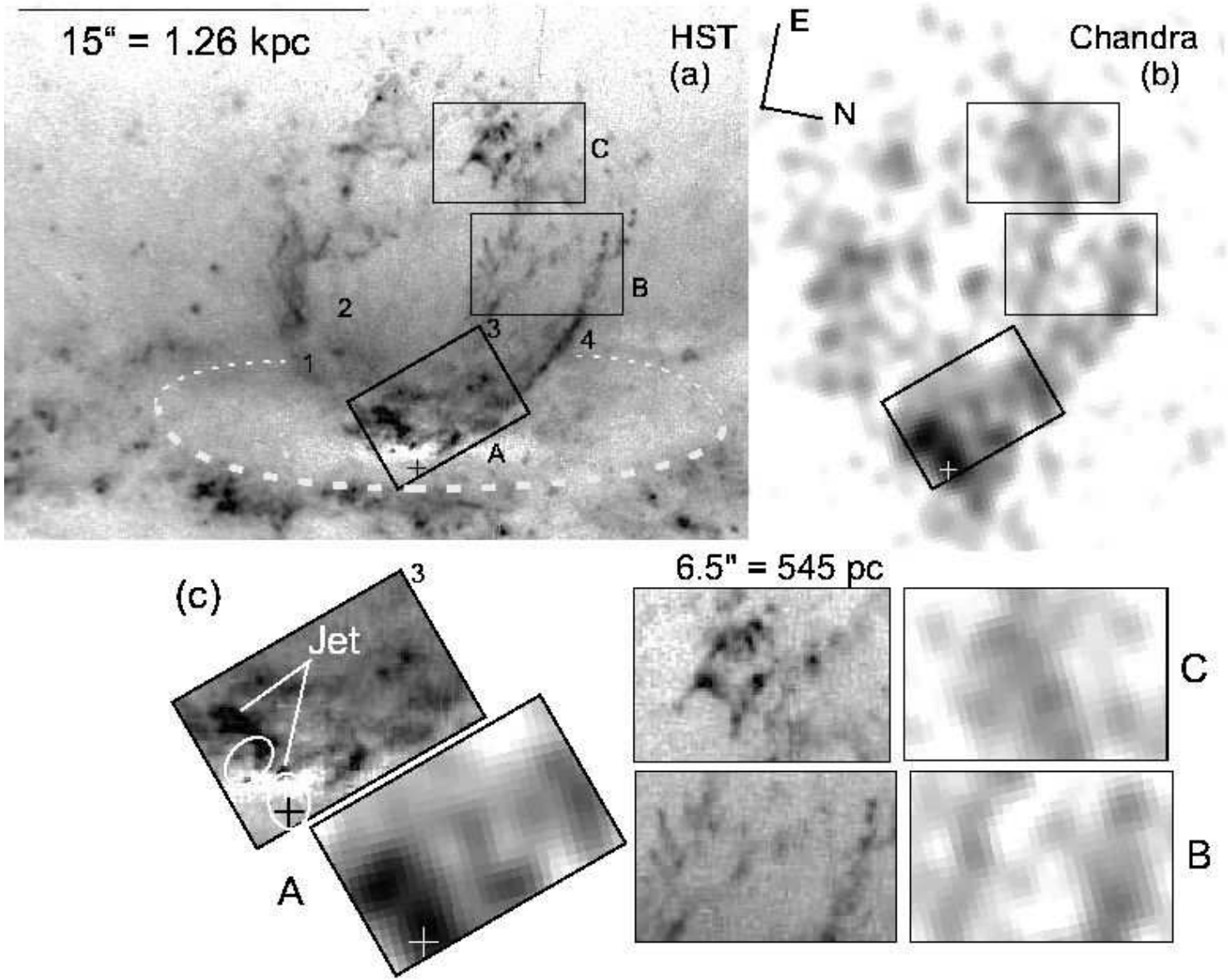}
\caption{\label{fig1}
(a) Dithered \emph{HST}/WFPC2 H\protect\( \alpha \protect \)+[\ion{N}{2}]
\protect\( \lambda \protect \)6583 minus I-band image of the
superbubble. The ``+'' spans the current uncertainty
in the nucleus position. The bubble is formed from vertical towers
of filaments anchored to the galaxy disk, numbered \protect\( 1-4\protect \)
from south to north. The nuclear wind appears to have ``scoured''
disk gas and dust from within the white dashed circle. (b) 26.6 ksec
\emph{Chandra}/ACIS image at the same scale and orientation, smoothed
to 0\farcs8 FWHM. Some regions of correspondence \protect\( A-C\protect \)
are detailed in panel (c); each box is \protect\( 545\times 360\protect \)
pc (6\farcs5\( \times 4\farcs 3\protect \)). Ovals on the optical
image in A register X-ray features.}
\end{figure}

\begin{figure}
\includegraphics[scale=0.9]{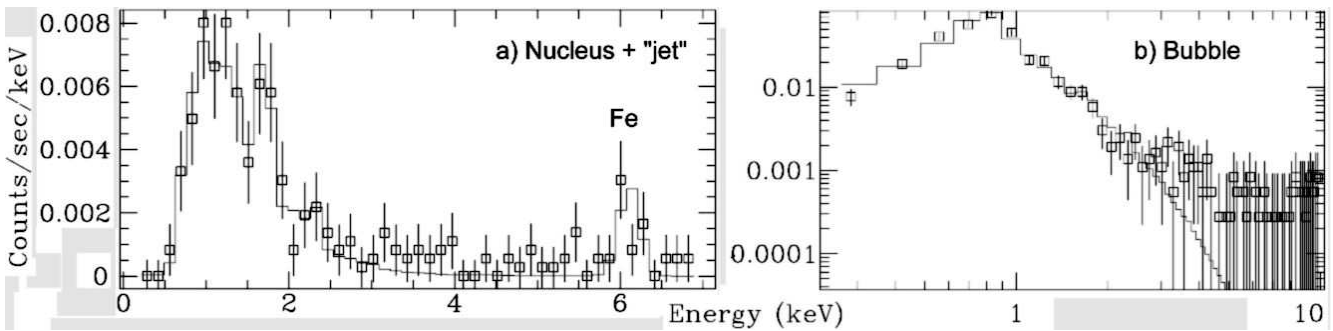}
\caption{\label{fig0} 
\emph{Chandra}/ACIS spectra extracted from elliptical regions (see text)
that are centered on
the a) nucleus+jet, and b) superbubble.  Solid lines show representative
fits assuming absorbed, single-temperature models; the nucleus+jet
spectrum also uses a single Gaussian to fit
the Fe K$\alpha$ line complex at 6 keV.
Fit parameters and uncertainties are summarized in Table 1.}
\end{figure}

Fig. \ref{fig1} shows that, within the superbubble, optical line
and X-ray filament networks coincide within our registration uncertainty
(\( \la 0\farcs 7 \) or 60 pc). In both wavebands, the two northernmost
towers (numbers 3 and 4 in Fig. \ref{fig1}a) are
less diffuse than the southern pair. All towers extend down to the
galaxy nuclear region.
CBVF show (see also Fig.\ 3) that radio emission from the eastern
superbubble peaks 
at larger galactic radii than the optical line and X-ray emission,
arguing against an inverse Compton origin for the X-rays.

Because of the similar appearance of the superbubble
in the two wavebands, we bounded the volume of warm gas
by assuming that it envelopes the optical-line emitting filaments.
The areas of the optical filaments are often upper limits, 
constrained by our inability to resolve many of them in the
[N~II]+H$\alpha$ line \emph{HST/WFPC2} image.  
From Table 3 of CBVF we summed the volume of those that we now
see are associated with X-rays, then increased this by a conservative factor of
16 in cross-sectional area (assuming cylindrical filaments)
to reach the resolution
of the \emph{Chandra} observations, yielding total emitting volume
$V\la 10^8$ pc$^3$.
We used the luminosities derived from our X-ray spectral fits to constrain
the average gas densities and other physical properties 
of the brightest parts of the superbubble; given the small number of X-ray
counts, the derived physical parameters
are at best order of magnitude estimates.
The luminosity of the X-ray emitting gas
$$L(T) = V_x n_{ex}n_{px}\Lambda(T)~\mathrm{erg~s^{-1}},$$
where $n_{ex}(n_{px})$ is the electron (proton)
density of the hot gas, and $\Lambda(T)$ is the cooling function 
that is approximated near $T_6 = 10^6$ K by
$$\Lambda(T) = 1.6\times 10^{-22} T_6^{-0.7} + 2.3\times 10^{-24} \sqrt{T_6}~\mathrm{erg~cm^3~s^{-1}}$$
\citep{mc87}.
Following \citet{st02}, we scale by factor $\mathcal{R}=0.15/Z_\mathrm{true}$
to show the dependence on abundances; in the following 
$\beta = \sqrt{\eta_x \mathcal{R}}$.
The X-ray electron density is 
$n_{ex} = \sqrt{\mathrm{EI}~\mathcal{R}/V\eta_x} = 18~\beta/\eta_x$ cm$^{-3}$, 
while the thermal pressure is $P_\mathrm{TH}/k = 1.7\times10^8 \beta/\eta_x$ K cm$^{-3}$. 
The mass of the warm X-ray gas is $M_x = 4.5\times10^7 \beta~\mathrm{M_\odot}$. 
The time to cool to 10$^4$ K at solar abundance is
$t_c \approx 3.2\times10^3/n_x T^{3/2}$ s = $0.15 \beta$ Myr \citep{su93}.
The mass flow rate $\dot{M}_\mathrm{flow} \sim M_x v /z = 296 \beta v_{1000}~\mathrm{M_\odot ~yr^{-1}}$, 
where $z=1200$ pc is the height of the bubble, and $v_{1000}$ is the 
mass weighted, average outflow
velocity in units of $10^3$ km~s$^{-1}$ (typical of measured values for the
optical filaments, CBVF).
The soft X-ray plasma has thermal and kinetic energies
$E_\mathrm{TH} = 2\times 10^{56} \beta$ erg and $E_\mathrm{KE} = 5\times10^{54} \beta v_{1000}^2$ erg.
In comparison, we found (CBVF) that the optical-line emitting gas has
$E_\mathrm{TH} = 3\times 10^{54}/n_{ev}$ erg and
$E_\mathrm{KE} = 2\times 10^{54} \eta_v^{0.5} v_{1000}^2$ erg where $\eta_v$ 
and $n_{ev}$ are the filling factor and electron density of that gas phase.

The mass of both the visual line- and X-ray emitting gas can be written as 
$X(1~\mathrm{cm}^{-3}/n_e)~M_\odot$, because the emissivity in both cases is 
proportional to $n_p n_e$. From the \emph{HST} line image, we derived
(Table 3 of CBVF) that
$n_{ev} \ge 4.3 \eta_v^{-0.5}$ and $M_v = 1.3\times 10^6 \eta_v^{0.5}$ M$_\odot$.
If we assume that the two phases are in local pressure
equilibrium, we find the \emph{relative} values of electron density 
$(n_{ev}/n_{ex}=0.2)$, and mass $(M_v/M_x=0.03)$.

\subsection{X-rays From the Radio Counter-bubble}

Fig.\ \ref{fig2} maps X-rays over a larger scale and compares to the
radio continuum. A pair of faint ridges of X-rays emerge from the
nucleus and trend westward, toward the bottom. They have the same
opening angle as the bottom half of the bubble walls (forming a cone),
suggesting that they trace the walls of the base of the radio counterbubble. If
of similar surface brightness to the eastern bubble, they are attenuated
about threefold in count rate by absorption in the galaxy, so are
too faint for meaningful spectral analysis.

\subsection{X-ray Emission Near the Disk, Outside the Superbubble}

\begin{figure}
\includegraphics{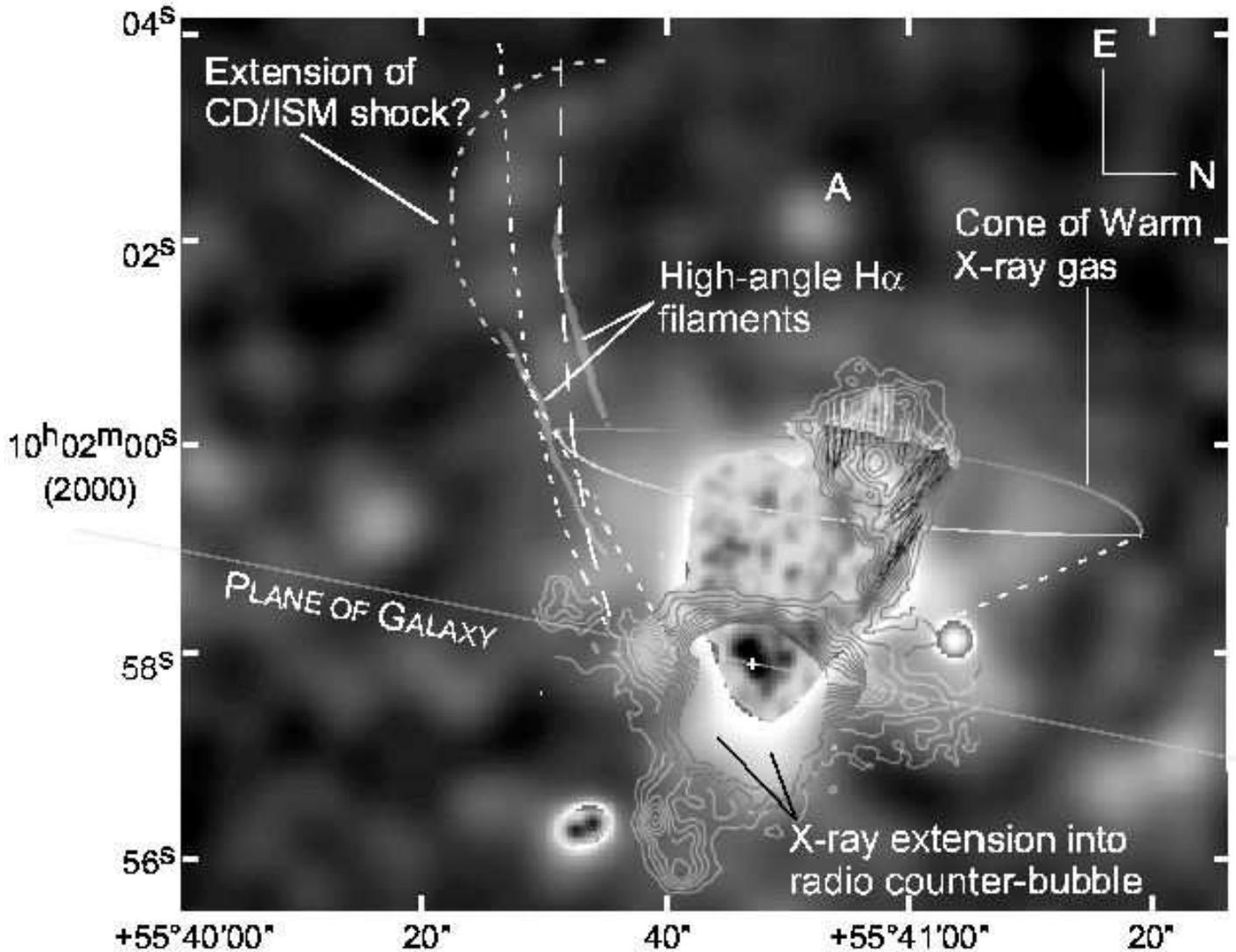}
\caption{\label{fig2}
Large scale X-ray grayscale and radio contours (3.8 cm VLA image from CBVF)
are compared.
The grayscale is smoothed
to 2\arcsec\ FWHM until it reverses to show X-ray filaments with 1\arcsec\ smoothing.
Contours, suppressed near and immediately west of the nucleus, are
drawn at (0.6, 0.55, 0.5, 0.45, 0.4, 0.35, 0.3, 0.25, 0.2,
0.15, 0.1, 0.05) percent of the peak brightness of 96.2 mJy
beam\protect\( ^{-1}\protect \) (HPBW 1\farcs6); the lowest two are
omitted east of the nucleus to minimize clutter. Also shown are 
the magnetic field vectors. The almost vertical
white dotted and dashed lines from CBVF show the modeled positions
of the contact discontinuity/ISM shock for power-law and exponentially
declining gas density above the galaxy disk. They coincide with regions
of diffuse X-ray emission and with several high-angle H\protect\( \alpha \protect \)
filaments shown in gray.}
\end{figure}

Fig.\ \ref{fig2} shows that diffuse X-ray emission surrounds the nucleus
out to 25\arcsec\ radius. Emission is more extensive east of the nucleus,
the disk nearside, than to the west. X-rays fill a cone with
opening angle $\sim$90\( ^{\circ } \) as height increases above the disk, and
are more concentrated on the south side of the nucleus where the cone
edge coincides with inclined linear H\( \alpha  \) filaments. X-ray
emission on the north side of the nucleus is patchier, but may also
be bounded in the cone sketched in Fig. \ref{fig2}. The cone apex
is not at the galaxy nucleus. Instead, the cone reaches the disk while
still open, then becomes ``two horned'' west of the nucleus as
X-rays trace the radio counterbubble (discussed previously).  
Again, X-ray emission is too faint for meaningful spectral constraints.

\subsection{X-ray Emission Above the Galaxy Disk}

We have not analyzed the statistical significance of other X-ray point sources.
However, one of the registration sources
in both wavebands is a small galaxy in the \emph{HST} I-band
image 32\arcsec\ east of the nucleus (source ``A'' in Fig. \ref{fig2}).
An arc of 4 sources, each about half the X-ray brightness of this 
galaxy but among the brighter ``background'' sources, 
lies \( \sim  \)20\arcsec\ south and may form the eastern extension
of the southern edge of the X-ray cone. It curves toward the cone
axis as height increases from 30 to 40\arcsec.

\section{Discussion}

\subsection{Outer Wind/ISM Shock}

Proceeding inward from the largest scale, the region of diffuse X-ray
emission is bounded by the optically emitting X-pattern of filaments,
which CBVF showed have the shape expected \citep{sc86} for the contact
discontinuity/ISM shock associated with lateral stagnation of the
superwind in the galaxy {}``thick disk''/halo; the X-ray emission is not
significantly edge brightened, suggesting a partially filled volume of warm
gas within the shocked wind, not a shell of conductively heated gas.
CBVF's models appropriate for steady-state wind flow, and reproduced
as dotted and dashed white lines in Fig. \ref{fig2}, remain open
at the top. However, \citet{so01} have modeled an adiabatic ``bipolar
hypershell'' that arises from a point, impulsive energy release on
a timescale much shorter than the several Myr expansion time of
the superbubble. These structures close at the top. The appearance of
their models is sensitive to the density distributions of disk, halo,
and the density of the extragalactic medium, but resemble one another
for roughly the first 10 Myrs. The outer shell expands at 200 -- 300
km~s\( ^{-1} \) to shock-heat gas to 0.2 keV (so would be very {}``soft''
in a {}``hardness'' map derived from a deeper \emph{Chandra} exposure),
consistent with the velocities in our Fabry-Perot spectra \citep{ve94}.
Of course, identical shock envelopes arise for constant values of
\( E_{0}/\rho _{0} \), where \( E_{0} \) is the energy of the explosive
event and \( \rho _{0} \) the ambient gas density. \citet{st02} considered
this model to interpret their \emph{Chandra} image of the starburst
galaxy NGC 253, rejecting it for one that describes optical
emission from a wind-blown shell \citep{we77}, the model that 
\citep{ve94} have applied to NGC 3079.

\citet{so01} compared their models to PTV's PSPC image of NGC 3079,
noting double-horned features on both sides of the galaxy at radii
of $\sim$100\arcsec. This is much larger than the scale we are discussing,
suggesting either that there have been several outbursts over the
last 100 Myrs or that the larger scale double-horns seen by the \emph{ROSAT}
PSPC are actually spurious overlaps of unrelated background objects.

As discussed in \S\ref{sec:towers}, what powers the superbubble 
remains ambiguous despite detection of Fe line emission (Fig.\ \ref{fig0}).
As emphasized by \citet{ma99}, it is extremely difficult to distinguish
the action of a starburst from an AGN based solely on the ionizing flux.
Deeper images with \emph{Chandra} and especially \emph{XMM/Newton}
could establish if the contact discontinuity/ISM shock
closes to form symmetric bubbles. Open morphology would argue for
a steady wind rather than impulsive starburst, supporting the interpretation
of \citet{ir87} that \ion{H}{1} in the companion galaxy 
NGC 3073, 10\arcmin\ west,
trails away from NGC 3079 because of what would have to be an extremely
energetic wind. A steady
wind would favor an AGN power source, because repetitive
starbursts would make numerous, less massive stars that are not evident
\citep[cf.\ those in M82,][]{dg01}.

\subsection{Filament Correlation Within the Superbubble}

The organization of both filament systems into vertical towers attached
to the galaxy disk excludes the possibility that the optical filaments
are halo clouds caught in the wind. For the superwind in NGC 253,
\citet{st02} dismiss the possibility that diffuse X-ray and filamentary
H\( \alpha  \)
emission both come from a cooling, volume filling wind; their argument
is also valid for NGC 3079. 

That the towers emit X-rays along their
\emph{entire} optical length is an important constraint, because the radial
velocities of the optically emitting filaments show a linear gradient
from 200 to $>$1000 km~s\( ^{-1} \) with height above the disk \citep{ve94}
until they overturn in a vortex in the top third of the superbubble
(CBVF). The ubiquitous X-ray emission
suggests that the superwind has driven cool disk gas
into the halo, with X-rays being emitted either as upstream, stand-off
bow shocks or by cooling at cloud/wind conductive interfaces. In either
case, the X-ray emitting skin is thin and would plausibly 
coincide at our resolution
with the optical line emitting filaments, e.g.\ for preshock density
$n_0$ cm$^{-3}$, 
the separation would be $1\times V_{s500}^4/n_0$ arcsec scaled for shock
speed in units of 500 km~s$^{-1}$. In this picture, the wind
strips gas from the walls of the dense disk cavity, and this gas then
fragments into disintegrating streams by Kelvin-Helmholtz instabilities.
The fragments gradually disperse, and eventually become Rayleigh-Taylor
unstable in the top third of the superbubble where they pass through
the vortex and begin to drip toward the disk. The filaments are embedded
in the wind that flows past them to reach into the halo.

The soft X-rays may arise as optical line-emitting clouds evaporate into a
region of unseen, hot (10$^8$ K) X-ray gas that fills the superbubble.
Because the cloud column density is surely $< 10^{24}$ cm$^{-2}$, we can use 
the saturated 
thermal conductive rates determined by \citet{kr81}.  The approximate cloud 
lifetime for saturated thermal evaporation without an inhibiting magnetic
field is
$$\tau_\mathrm{TE} \approx 0.3~\mathrm{Myr}~N_{23}^{7/6}T_{4}^{1/6}/P_9,$$
where $P_9$ is the thermal cloud pressure in $10^9$ 
erg~cm$^{-3}$ ($nT=10^7$ K cm$^{-3}$),
$T_4$ is the cloud temperature in $10^4$ K, and $N_{23}$ is the
cloud column density in 10$^{23}$ cm$^{-2}$ that is unfortunately unconstrained
by existing optical spectra of the superbubble. However, because clouds reach
high into the galaxy halo from the disk, we can set a lower limit on
their column densities by equating $\tau_\mathrm{TE}$
to the cloud crossing time $z/v_{1000} \approx 1.2$ Myr, obtaining
$N_{23} \approx 0.8 T_c^{-1/7}P_9^{6/7}$.
The visible clouds that have survived conductive evaporation are 
therefore those with column densities $>10^{22}$ cm$^{-3}$ and hence large sizes
(if cylindrical, $r > 0.3$ pc) and total masses $M > 10$ M$_\odot$, 
consistent with the ionized masses determined in CBVF.

\subsection{Wind Energetics}

The detection of Fe K line emission addresses the question of whether the AGN is
directly responsible for the superbubble.
Extended filamentation is seen in both
Seyfert galaxies (where circumnuclear starbursts are commonly observed), and
in starburst galaxies that do not show strong nuclear X-ray sources.
Starburst galaxies commonly show an ionization gradient in the extended 
filaments, in the sense that inner filaments appear to be photoionized while
the outer ones are more shock-like \citep[for example]{ma97,sh98}, although
\citet{ve02} have found a starburst-driven filament 
system that is shock-like throughout. 
From morphology alone, it is difficult to distinguish a shock from a 
photoionization process: hydrodynamics can shape filaments that are then
photoionized by the central starburst.
In the case of M82, a well defined ionization cone is mapped in e.g.\
[\ion{N}{2}]/H$\alpha$ \citep{sh98}). The optical cone appears
to be mostly photoionized from stars whereas the X-ray cone appears to be
shock excited. In NGC 1068, where winds and radiation are both known to be
important, the AGN radiation field is thought to explain both the optical
and X-ray emission, and this object appears to have an X-ray photoionization
cone \citep{yo01}.
On UV energetics alone, it is also impossible to distinguish between AGN or
starburst \citep[for example]{ma99}.
A gas temperature derived from both the X-ray and the optical emission is
crucial to establishing the energy source. The mean energy of an ionizing photon
in photoionized gas is much smaller than that required to ionize the medium
collisionally, hence shock heated gas is considerably hotter.
Suitable optical data do not yet exist for the superbubble of NGC 3079.

True wind energetics are elusive because they depend on both the supernova
source energetics and the thermalization efficiency $\varepsilon$. 
Before \emph{Chandra}, X-ray observations of starburst cores had lacked 
spatial resolution to resolve the contributions of X-ray 
binaries. Now, ACIS on \emph{Chandra}
can search for hot gas in the starburst core, and \citet{gr00} argue
that they have detected it in M82.
Summing the energy in a wind like that in M82 with best estimates
for the filling factors in various gas phases, one finds that it is still 
substantially below the energy derived using the radio-determined supernova
rate and a moderate to high (20--50\%) thermalization
efficiency. Moreover, \citet{ch01} argue that the supernova rate in M82 may
be substantially higher than inferred from radio counts. These suggest
that we are missing much of the outflow energy, in agreement with multi-phase
hydrodynamical models \citep{su94,st00}.

The problem is that the outflowing gas is almost all entrained disk and halo 
gas, not wind material.
As evidence for this, the outer filaments in M82 \citep[and NGC 1482,][]{ve02}
share the rotation of the galaxy disk, and the
pattern of magnetic fields and gas motions in NGC 3079 (CBVF) strongly
suggest that the top of the superbubble is forming a vortex. Second, hot gas 
exists in superwinds \citep{he01} and appears to be moving faster than the 
cooler gas, inconsistent with standard superbubble models \citep{we77} but
resembling the pattern seen in hydrodynamical wind simulations \citep{su94,st00}.

Evidence for extensive supernova remnants in the disk of NGC 3079 is ambiguous
(see \S4.3 of CBVF).
We thus address wind energetics from the constraints derived previously on 
the efficiency of thermalization.
The energetics of the soft X-ray component in the superbubble that
we constrained in \S\ref{sec:towers} are comparable to those
determined by \citet{st02} 
for the diffuse wind in NGC 253. This arises even though the X-ray emitting
filaments in NGC 3079 may have much higher density.

\section{Conclusions}

\emph{Chandra} reveals a remarkably
tight correspondence ($<$0\farcs8, 65 pc) between
optical line and X-ray emitting filaments in the superbubble of NGC
3079. The bubble is embedded in a conical distribution of more diffuse
X-rays. West of the nucleus, this emission extends into the optically
obscured, radio counterbubble. East of the nucleus (i.e.\ on the near
side of the galaxy) a funnel is also delineated by extremely narrow
optical line emitting filaments that rise several kpc above the galaxy
disk, supporting our earlier contention that they form the contact
discontinuity/halo shock of the wind. These filaments connect back
to the disk at one end and at the other, perhaps, to an X-ray ridge
that seems to curve back toward the cone axis. However, at present
X-ray sensitivity, the loop does not close completely, so the wind
may escape the galaxy. A deeper \emph{Chandra} exposure would confirm 
this X-ray extension, and would
double spatial resolution of the superbubble to tighten the volume
of the emission integral hence constraints on derived physical parameters.

Even without a complete understanding of the physical process,
the close association between the \emph{Chandra} and \emph{HST} 
emission is important for several reasons. First, powerful winds 
must be involved either directly or indirectly in producing this 
filamentary complex. Second, high resolution joint \emph{Chandra/HST} studies
of the filaments in this galaxy, M82 \citep{gr00}, NGC 253 \citep{st00a,st02}, 
and NGC 1068 \citep{yo01}
show that both the X-ray and optical emission is highly clumped, with
resolved structures seen down to our resolutions of 65 pc (\emph{Chandra})
and 15 pc (\emph{HST}).
The evidence for entrainment in these winds is compelling
(e.g.\ CBVF), and we suspect that most of the superwind mass resides 
in the filamentary complex.  We agree with the assessment of \citet{st01}
that the smallest cloud scales in simulations must be very
well sampled to treat cloud entrainment and acceleration
correctly \citep{kmc94}.  The resolution of present wind
models \citep{su94,st00} is ten times too large to match
the observed clumping scales,
arguing for a new generation of hydrodynamical models to
establish testable predictions on cloud internal properties.
In the meantime,
we may be seriously underestimating the mass lost in superwinds, and 
therefore their influence within and around the host galaxy.

\acknowledgements{}

We thank the referee for useful comments.
Our research with the \emph{CXO} is supported by GTO grant 01700250.
S.V. acknowledges partial support of this research by a Cottrell
Scholarship awarded by the Research Corporation, NASA/LTSA grant NAG
56547, and NSF/CAREER grant AST-9874973.

\begin{deluxetable}{lcccccr}
\tabletypesize{\small}
\tablecaption{X-ray Components in the Circumnuclear Region of NGC 3079}
\tablewidth{0pc}
\tablehead{\colhead{Spectral} & \colhead{n$_H$ \tablenotemark{a}} & \colhead{kT} & \colhead{Abundance} & \colhead{EI\tablenotemark{b}} & \colhead{Flux\tablenotemark{c}} & \colhead{Luminosity\tablenotemark{d}} \\
\colhead{region} & \colhead{($\times 10^{22}~\mathrm{atoms~cm^{-2}}$)} & \colhead{(keV)} & \colhead{(relative to solar)} & \colhead{cm$^{-3}$} & \colhead{(erg~s$^{-1}$cm$^{-2}$)} & \colhead{(erg)} }
\startdata
Nucleus+jet & $1.7\pm0.2$ & 0.55$\pm0.05$ & 0.52 &$2.2\pm0.7\times 10^{66}$ & $3\times10^{-11}$ & $1.1\pm0.1\times 10^{42}$ \\
Fe line &\nodata &\nodata &\nodata & \nodata &9$\times10^{-12}$ & $3.3\pm0.5\times10^{41}$ \\
Bubble\tablenotemark{e} & 0.3 & 0.8 & 0.16 & 10$^{66}$ & $8\times10^{-11}$ & $3\times10^{42}$ \\
\tablenotetext{a}{Photoelectric absorption using the formulation of
\citet{mm83}.}
\tablenotetext{b}{Model component normalizations, 
where the fitted emission ``integral" 
$EI = n_e n_H \eta_x V$ and $\eta_X$ is the filling factor of volume $V$.}
\tablenotetext{c}{Integrated across passband 0.1--6.5 keV.}
\tablenotetext{d}{Distance 17.3 Mpc and isotropic emission.}
\tablenotetext{e}{The lack of distinct features in this spectrum
prohibits error estimates for this component, and formal $\chi^2$ results
in general for all components. Where reported, ranges on derived parameters
span those values that produced nearly identical ``reasonable'' fits.}
\enddata
\end{deluxetable}

\begin{thebibliography}{}
\bibitem[Anders \& Grevesse(1989)]{an89}Anders, E. \& Grevesse, N. 1989, \gca,
53, 197
\bibitem[Carlbeg(1999)]{ca99}Carlberg, R. 1999, in The Formation of Galactic
Bulges, eds. C. M. Carollo, H. C. Ferguson, \& R. F. G. Wyse (New York: CUP), 64
\bibitem[Chevalier \& Clegg(1985)]{cc85}Chevalier, R. A. \& Clegg, A. W. 1985, \nat, 317, 44
\bibitem[Cecil et al.(2001)]{ce01}Cecil, G., Bland-Hawthorn, J., Veilleux, S.,
\& Filippenko, A. V. 2001, \apj, 555, 338 (CBVF)
\bibitem[Chevalier \& Fransson(2001)]{ch01}Chevalier, R. A. \& Fransson, C. 2001,
\apj, 558, L27
\bibitem[de Grijs, O'Connell, \& Gallagher(2001)]{dg01}de Grijs, R., O'Connell, R. W.,
\& Gallagher, J. S. 2001, \aj, 121, 768
\bibitem[Duric \& Seaquist(1988)]{du88}Duric, N. \& Seaquist, R. S. 1988, \apj, 326, 574
\bibitem[Efstathiou(2000)]{ef00}Efstathiou, G. 2000, \mnras, 317, 697
\bibitem[Ford et al.(1986)]{fo86}Ford, H. C., Dahari, O., Jacoby, G. H., Crane,
P. C., \& Ciardullo, R. 1986, \apj, 311, L7

\bibitem[Freeman \& Bland-Hawthorn(2002)]{bf02}Freeman, K. \& Bland-Hawthorn, J. 2002, \araa, 40, in press
\bibitem[Frye, Broadhurst, \& Benitez(2002)]{fr02}Frye, B., Broadhurst, T, \&
Benitez, N. 2002, \apj, 568, 558
\bibitem[Griffiths et al.(2000)]{gr00}Griffiths, R. E. et al. 2000, Science, 290,
1325
\bibitem[Hawarden et al.(1995)]{ha95}Hawarden, T. G. et al. 1995, \mnras, 276, 1197
\bibitem[Heckman et al.(2001)]{he01}Heckman, T. M. et al. 2001, \apj, 558, 56
\bibitem[Heckman(2002)]{he02}Heckman, T. M. 2002, in Extragalactic Gas at Low
Redshift, eds. J. Mulchaey \& J. Stocke, ASP Conf. Ser. 254, 292 (San Francisco: ASP)
\bibitem[Hollenbach \& McKee(1979)]{hm79}Hollenbach, P. J. \& McKee, C. F. 1979,
\apjs, 41, 555
\bibitem[Irwin et al.(1987)]{ir87}Irwin, J. A., Seaquist, E. R., Taylor, A. R., \& Duric, N. 1987, \apj,
313, L91
\bibitem[Klein, McKee, \& Colella(1994)]{kmc94}Klein, R. I., McKee, C. F., \& Colella, P. 1994, \apj, 420, 213
\bibitem[Krolik, McKee, \& Tarter(1981)]{kr81}Krolik, J., McKee, C. F., \& Tarter, 
C. B. 1981, \apj, 249, 422
\bibitem[Maloney(1999)]{ma99}Maloney, P. R. 1999, \apss, 266, 207
\bibitem[Martin(1997)]{ma97}Martin, C. 1997, \apj, 491, 561
\bibitem[McCray(1987)]{mc87}McCray, R. 1987, in Spectroscopy of Astrophysical
Plasmas, ed. A. Dalgarno \& D. Layzer (New York: Cambridge Univ. Press), chap.\ 10
\bibitem[Morrison \& McCammon(1983)]{mm83}Morrison, R. \& McCammon, D. 1983, \apj,
270, 119
\bibitem[Pietsch, Trinchieri, \& Vogler(1998)]{pi98}Pietsch, Trinchieri, \& Vogler 1998, \aap, 340, 351 (PTV)
\bibitem[Ptak(1997)]{pt97}Ptak, A. 1997, Ph. D thesis, Univ. Maryland, College Park
\bibitem[Schiano(1986)]{sc86}Schiano, A. V. R. 1986, \apj, 302, 81
\bibitem[Shopbell \& Bland-Hawthorn(1998)]{sh98}Shopbell, P. \& Bland-Hawthorn, J. 1998, \apj, 493, 129
\bibitem[Sofue et al.(2001)]{so01a}Sofue, Y. et al.\ 2001, \apjl, 547, L115
\bibitem[Sofue \& Vogler(2001)]{so01}Sofue, Y. \& Vogler, A. 2001, \aap, 370, 53
\bibitem[Strickland(2001)]{st01}Strickland, D. K. 2001, in Chemical Enrichment
of the ICM and IGM, ASP Conf. Ser., astro-ph/0107116
\bibitem[Strickland \& Stevens(2000)]{st00}Strickland, D. K. \& Stevens, I. R. 2002, \mnras, 314, 511
\bibitem[Strickland, Heckman, Weaver, \& Dahlem(2000)]{st00a}Strickland, D. K., 
Heckman, T. M., Weaver, K. A., \& Dahlem, M. 2000, \aj, 120, 2965
\bibitem[Strickland et al.(2002)]{st02}Strickland, D. K. et al. 2002, \apj, 568, 689
\bibitem[Suchkov et al.(1994)]{su94}Suchkov, A. A., Balsara, D. S., Heckman, T.
M., \& Leitherer, C. 1994, \apj, 430, 511
\bibitem[Sutherland \& Dopita(1993)]{su93}Sutherland, R. S. \& Dopita, M. A. 1993,
\apjs, 88, 253
\bibitem[Thornton et al.(1998)]{th98}Thornton, K., Gaudlitz, M., Janka, H.-Th., \&
Steinmetz, M. 1998, \apj, 500, 95
\bibitem[Tripp et al.(2002)]{tr02}Tripp, T. et al. 2002, \apj, astro-ph/0204204
\bibitem[Veilleux \& Rupke(2002)]{ve02}Veilleux, S. \& Rupke, D. S. 2002, \apj, 565, L63
\bibitem[Veilleux et al.(1994)]{ve94}Veilleux, S., Cecil, G., Bland-Hawthorn, J., Tully, R. B., Filippenko,
A. V., \& Sargent, W. L. W.1994, \apj, 433, 48
\bibitem[Weaver et al.(1977)]{we77}Weaver, R., McCray, R., Castor, J., Shapiro, P., \& Moore, R. 1977,
\apj, 218, 377
\bibitem[Young, Wilson, \& Shopbell(2001)]{yo01}Young, A. J., Wilson, A. S., 
\& Shopbell, P. L. S. 2001, \apj, 556, 6
\end{thebibliography}
\end{document}